\documentclass[twocolumn,english,aps,showpacs]{revtex4}
\usepackage[T1]{fontenc}
\usepackage[latin9]{inputenc}
\usepackage{color}
\usepackage{babel}

\usepackage{amsmath}
\usepackage{amssymb}
\usepackage[unicode=true, 
 bookmarks=true,bookmarksnumbered=false,bookmarksopen=false,
 breaklinks=false,pdfborder={0 0 0},backref=false,colorlinks=true]
 {hyperref}
\hypersetup{pdftitle={Unified model of k-inflation, dark matter & dark energy},
 pdfauthor={Nilok Bose, A. S. Majumdar}}
 
\makeatletter
\@ifundefined{definecolor}
 {\usepackage{color}}{}


\makeatother

\begin{document}

\title{Unified Model of \textit{k}-Inflation, Dark Matter \& Dark Energy}

\author{Nilok Bose}

\email{nilok@bose.res.in}

\author{A. S. Majumdar}

\email{archan@bose.res.in}

\affiliation{S. N. Bose National Centre for Basic Sciences, Block JD, Sector III,
Salt Lake, Calcutta 700098, India}

\date{\today}
\begin{abstract}
We present a \textit{k}-essence model where a single scalar field
is responsible for the early expansion of the universe through the
process of \textit{k}-inflation and at appropriate subsequent stages
acts both as dark matter and dark energy. The Lagrangian contains
a potential for the scalar field as well as a non-canonical kinetic
term, and is of the form $F(X)V(\phi)$ which has been widely used
as a \textit{k}-essence Lagrangian. After the period of inflation
is over the model can be approximated as purely kinetic \textit{k}-essence,
generating dark matter and dark energy at late times. We show how
observational results are used to put constraints on the parameters
of this model. 
\end{abstract}

\pacs{98.80.-k, 98.80.Cq, 95.36.+x}

\maketitle

\section{Introduction}

Till date the nature of both dark matter and dark energy is largely
unknown and they constitute one of the biggest puzzles of modern cosmology.
The dynamics of the process driving the current acceleration of the
universe is still unclear but there exist a wide variety of approaches
that could theoretically account for this acceleration. The combination
of observations of high red-shift supernovae, CMBR and large scale
structure have categorized the current energy density of the universe
to consist of approximately 73\% dark energy, which drives the late
time acceleration of the universe, and approximately 23\% dark matter
which clusters and is responsible for the formation of large-scale
structure in the universe (see \cite{Sahni} and references therein).
These observations, including those of the nearly scale-independent
density perturbations, are also in conformity with the widely held
view that the early universe underwent a brief period of accelerated
expansion, dubbed as inflation.

Since accelerated expansion is a common feature for both the very
early and the very late universe, it is plausible that some common
mechanism could be responsible for both. Several models have been
constructed to explain inflation and dark energy using a single scalar
field (see, for example, quintessential inflation \cite{Peebles}).
It is also possible for the two dark components of the universe to
be the manifestations of a single entity, and a considerable number
of models can be found in the literature that try to unify dark matter
and dark energy (for instance \cite{Scherrer}, \cite{Arbey}). Apart
from the above schemes there are models that try to unify inflation
and dark matter (for instance \cite{Lidsey}) and also those that
attempt to unify all three, \textit{viz}. inflation, dark matter and
dark energy (for instance \cite{triple}).

In many of these unification models the dynamics of one or more scalar
fields plays the central role. In fact, the idea of \textit{k}-essence
driven by scalar field with a non-canonical kinetic term motivated
from the Born-Infeld action of string theory \cite{born-infeld},
was first introduced as a possible model for inflation \cite{picon}.
Later, it was noted that \textit{k}-essence could also yield interesting
models for the dark energy \cite{chiba}, \cite{chimento}. An interesting attempt was
made to unify dark matter and dark energy using kinetic \textit{k}-essence
in \cite{Scherrer}. Though this model had its share of problems (it
is worth noting that a purely kinetic \textit{k}-essence leads to
a static universe when the late time energy density of the universe
is expressed simply as a sum of a cosmological constant and a dark
matter term \cite{Bose}), extensions of the formalism to extract
out dark matter and dark energy components within a unified framework
have been used also in subsequent works \cite{chimento3}.

Recently, we \cite{Bose} have proposed a \textit{k}-essence model
that reproduces the essential features of inflation, dark matter and
dark energy within a unified framework. We found that a couple of
parameters of this model had to be tuned in order to conform with
various observational features pertaining to both the early and the
late time eras of the universe. The Lagrangian chosen in this model
was of the form where the kinetic and potential terms were decoupled
in the standard way. However, it may be recalled that in most \textit{k}-essence
models \cite{chiba}, \cite{chimento} including the original \textit{k}-inflation idea \cite{picon},
the distinguishing feature was the use of non-canonical kinetic terms
in the Lagrangian of the form $F(X)V(\phi)$. In the present paper
we return to such a Lagrangian with the motivation of reproducing
the features of inflation in the early universe, and also generating
dark matter and dark energy at late times. We find that after the
early expansion is over, our present model can be approximated as
kinetic \textit{k}-essence, i.e., the dynamics becomes dominated by
only the kinetic component of the scalar field. We show that the late
time energy density reproduces a cosmological constant and a matter
like term which we call dark matter. We then consider observational
results from the both the early and late eras, which are used to put
constraints on the parameters of this model.

\section{The Model}

We begin with a Lagrangian for a scalar field $\phi$ of the form\begin{equation}
\mathcal{L}=F(X)V(\phi)\label{eq:1}\end{equation}
 where $X$ is defined as\[
X=\dfrac{1}{2}\partial_{\mu}\phi\partial^{\mu}\phi\]
Throughout this paper we will work with a flat Robertson-Walker metric
having signature $(+,-,-,-)$. Taking the scalar field to be homogeneous
in space, which is the usual case, we get $X=\dfrac{1}{2}\dot{\phi^{2}}$.

The functional forms of \textit{F} and \textit{V} are taken to be\begin{equation}
F(X)=KX-m_{Pl}^{2}L\sqrt{X}+m_{Pl}^{4}M\label{eq:2}\end{equation}
 \begin{equation}
V(\phi)=1+e^{-\phi/\phi_{c}}\label{eq:3}\end{equation}
 where the parameters $K$, $L$ and $M$ are dimensionless, and are
taken to be positive. The parameter $\phi_{c}$ is also taken to be
positive and clearly has the dimension of $\phi$. We work in natural
units and consider \textit{V} to be dimensionless. As is the usual
case, the scalar field $\phi$ has the dimension of mass. From the
definition of \textit{X} it turns out that \textit{X} and hence \textit{F}
has dimension $M^{4}$.

The energy density in this case is given by\begin{equation}
\rho=V(\phi)(2XF_{X}-F)\label{eq:4}\end{equation}
 where $F_{X}\equiv dF/dX$. So substituting the forms of \textit{F}
and \textit{V} in \eqref{eq:4} we get\begin{equation}
\rho=(1+e^{-\phi/\phi_{c}})(KX-m_{Pl}^{4}M)\label{eq:5}\end{equation}
 The pressure, which is simply the Lagrangian, turns out to be\begin{equation}
p=(1+e^{-\phi/\phi_{c}})(KX-m_{Pl}^{2}L\sqrt{X}+m_{Pl}^{4}M)\label{eq:6}\end{equation}
 The equation of state parameter is given by\begin{equation}
w=\dfrac{F}{2XF_{X}-F}\label{eq:9}\end{equation}
 which in our model evaluates to\begin{equation}
w=\dfrac{KX-L\sqrt{X}+M}{KX-M}\label{eq:10}\end{equation}
 The sound speed, or the speed at which perturbations travel,  is defined
to be  \cite{Garriga} \begin{equation}
c_{s}^{2}\equiv\dfrac{\partial p/\partial X}{\partial\rho/\partial X}=\dfrac{F_{X}}{2XF_{XX}+F_{X}}\label{eq:11}\end{equation}
 where $F_{XX}=d^{2}F/dX^{2}$. Note that this definition is different from
the usual definition of the adiabatic sound speed (namely, 
$c_{s}^{2}=\dfrac{dp}{d\rho}$). However, it has been shown recently \cite{sound}
that perturbations in such models travel with a speed defined as above, where
the authors also define this to be the ``phase speed''.

Now, the equation of motion for the \textit{k}-essence scalar field
is given by \begin{equation}
(2XF_{XX}+F_{X})\dot{X}+6HF_{X}X+\dfrac{\dot{V}}{V}(2XF_{X}-F)=0\label{eq:12}\end{equation}
 which has been written in terms of \textit{X}. If \textit{V} is a
constant or varies very slowly with time so that the third term in
\eqref{eq:12} is negligible then the situation corresponds to kinetic
\textit{k}-essence and the field equation can be written as\begin{equation}
(2XF_{XX}+F_{X})\dot{X}+6HF_{X}X=0\label{eq:13}\end{equation}
 This can be integrated exactly \cite{Scherrer} to give the solution\begin{equation}
\sqrt{X}F_{X}=\dfrac{k}{a^{3}}\label{eq:14}\end{equation}
 where \textit{k} is a constant of integration. This solution was previously
derived in a slightly different form in Ref. \cite{chimento}. 
The above result holds
irrespective of the spatial curvature of the universe.

The energy conservation equation states that \begin{equation}
\dot{\rho}=-3H(\rho+p)=-6HF_{X}XV\label{eq:15}\end{equation}
 This shows that the fixed points of the equation correspond to the
extrema of \textit{F} \cite{picon}, which from equations \eqref{eq:1}
and \eqref{eq:4} yields $\rho=-p$. Moreover $\rho$ decreases with
time when $\rho>-p$ and increases when $\rho<-p$ showing that any
point corresponding to $\rho=-p$ is an attractor and, as is well
known, will lead to exponential inflation.

In our model the extrema of \textit{F} correspond to $X=0$, or $X=m_{Pl}^{4}\dfrac{L^{2}}{4K^{2}}$.
The point $X=0$ is of no significance since that corresponds to energy
density and pressure which are constant in time. We take\begin{equation}
X_{0}=m_{Pl}^{4}\dfrac{L^{2}}{4K^{2}}\label{eq:16-1}\end{equation}
 which leads form definition of \textit{X}, to \begin{equation}
\dot{\phi}_{0}=m_{Pl}^{2}\dfrac{L}{\sqrt{2}K}\label{eq:16-2}\end{equation}
 where we have taken the positive sign for $\dot{\phi}$. For the
above value of \textit{X} the energy density and pressure turn out
to be \begin{equation}
\rho=V(\phi)\left(\dfrac{L^{2}}{4K}-M\right)m_{Pl}^{4}=-p\label{eq:16}\end{equation}

Actually, $X_{0}$ corresponds to an instantaneous attractive fixed
point and \textit{X} evolves slowly away from that point, which is
the analog of {}``slow-roll'' potential driven inflation in which
the potential dominates the kinetic term and evolves slowly. Hence,
in direct analogy, the above calculated values of $\rho$ and \textit{p}
can be called the {}``slow-roll'' values. In our model we assume
that the exponential term inside \textit{V} is much larger than $1$
during the course of inflation, for which
we must have $\phi_{0}/\phi_{c}<0$, and also $|\phi_{0}/\phi_{c}|\gg1$.
From Eq.(\ref{eq:15})we can
write $\phi_{0}=\dot{\phi_{0}}t+C_{\phi}$,
where $C_{\phi}$ is an integration constant. This constant can have
a negative value, hence making $\phi_{0}<0$. Thus, we choose $\phi_c > 0$,
such that the conditions $\phi_{0}/\phi_{c}<0$, and $|\phi_{0}/\phi_{c}|\gg1$ 
are satisfied during inflation. Since $\dot{\phi_{0}}>0$,  it follows that 
$\phi$ becomes less and less negative with time. 
\textit{V} can be quite accurately
approximated as $e^{-\phi/\phi_{c}}$. This enables us to find the
number of e-folds of expansion \textit{N}, under this {}``slow-roll''
approximation as \begin{equation}
N=\int\limits _{t_{i}}^{t_{e}}H\, dt=\int\limits _{\phi_{i}}^{\phi_{e}}H\,\dfrac{d\phi}{\dot{\phi}}\label{eq:16b}\end{equation}
 which turns out to be \begin{equation}
N\simeq\sqrt{\dfrac{8\pi}{3}}m_{Pl}^{-1}\left(\dfrac{L^{2}}{4K}-M\right)^{1/2}\dfrac{\sqrt{2}K}{L}2\phi_{c}\left(\sqrt{V}_{i}-\sqrt{V}_{e}\right)\label{eq:17}\end{equation}
 where the subscripts `i' and `e' refer to the intial and final values
respectively.

The slow-roll condition for \textit{k}-inflation is given by $[{\delta X}/{X_{0}}]\ll1$.
Now, during the post slow-roll stage we can write $X=X_{0}+\delta X$.
Also, from Eq.\eqref{eq:15}, one has \begin{equation}
\dfrac{F_{X}}{\left(KX-m_{Pl}^{4}M\right)}=-\dfrac{1}{6X}\dfrac{\dot{V}}{HV}\label{eq:17b}\end{equation}
 Retaining terms up to the first order in $\delta X$ we get \begin{equation}
\dfrac{\delta X}{X_{0}}\simeq\dfrac{\dfrac{1}{X_{0}}\left(\dfrac{L^{2}}{4K}-M\right)}{\sqrt{3\pi}\dfrac{L\phi_{c}}{X_{0}m_{Pl}^{2}}\left(\dfrac{L^{2}}{4K}-M\right)^{1/2}\sqrt{V}-\dfrac{K}{m_{Pl}^{4}}}\label{eq:18}\end{equation}
 Inflation ends when $\dfrac{\delta X}{X_{0}}\sim1$. Using this fact
in Eq.\eqref{eq:18} we can find the expression for the final value
of the potential, $V_{e}$ to be \begin{equation}
\begin{array}{c}
\sqrt{V_{e}}\simeq\dfrac{m_{Pl}}{\sqrt{3\pi}}\dfrac{1}{L\phi_{c}}\left(\dfrac{L^{2}}{4K}-M\right)^{1/2}\,\,\,\,\,\,\,\,\,\,\,\,\,\,\\
\\+\dfrac{m_{Pl}}{\sqrt{3\pi}}\dfrac{L}{4K\phi_{c}}\left(\dfrac{L^{2}}{4K}-M\right)^{-1/2}\end{array}\label{eq:19}\end{equation}

The kinematics of the inflationary era in our model may be viewed
in the following way. We start with some representative point in the
$\left(\rho,p\right)$ plane corresponding to some initial value of
$\phi$ such that the slow-roll condition is satisfied. In fact, during
the first evolutionary stage the representative point takes only a
few e-folds to reach the nearest inflationary attractor that corresponds
to $\rho=-p$ . After this initial stage the representative point
follows the post slow-roll motion, $X=X_{0}+\delta X$ with $\delta X/X_{0}\ll1$,
thereby staying near but not exactly on the $\rho=-p$ line. The value
of $\delta X$ is positive (as we will show later in the Section on
observational constraints). Hence \textit{X} slowly moves away from
the value $X_{0}$. As the evolution continues, the slow-roll condition
is satisfied to a less and lesser extent till a time is reached when
the slow-roll condition is actually violated ($\delta X/X_{0}\sim1$),
and one naturally exits the inflationary stage.

Now, after inflation ends we have $X>X_{0}$, meaning that the time evolution
of $\phi$ is faster than during inflation, and hence its value
increases very quickly and correspondingly decreases the value of the
exponential part in $V$, so that one gets $V\simeq1$.
In order for such a behaviour to ensue, we must have $\phi/\phi_{c}>0$
after inflation is over. Since we have already chosen $\phi_{c}$ to be
positive, then $\phi$ has to become positive after inflation where
previously it was negative, and this is exactly its behaviour as pointed
out earlier, i.e., $\dot{\phi_{0}}$ is positive.
Note that even if the ratio $\phi/\phi_c$ is not too big compared to 1,
the exponential part of the potential will be negligible. Thus,
after the inflationary expansion is over the exponential part in \textit{V}
quickly decays away (we will present an estimate of the time taken for this
process in the section on observational constraints on the model). When the
exponential term becomes quite negligible we have\[
V\approx1\,\,,\,\,\dot{V}\approx0\]
 So the field equation effectively becomes of the form of Eq.\eqref{eq:13}
and the dynamics can be approximated quite well by the purely kinetic
form of \textit{k}-essence. On using Eq.\eqref{eq:14} to find \textit{X}
as a function of \textit{a} we get \begin{equation}
X=\dfrac{1}{K^{2}}\left(m_{Pl}^{2}\dfrac{L}{2}+\dfrac{k}{a^{3}}\right)^{2}\label{eq:20}\end{equation}
 Therefore the corresponding expression for the \textit{k}-essence
energy density turns out to be \begin{equation}
\rho=m_{Pl}^{4}\left(\dfrac{L^{2}}{4K}-M\right)+m_{Pl}^{2}\dfrac{kL}{Ka^{3}}+\dfrac{k^{2}}{Ka^{6}}\label{eq:21}\end{equation}

The subsequent evolution of the universe is described as follows.
After the end of inflation the universe is in a kinetic dominated
period when the third term in Eq.\eqref{eq:21} dominates, which corresponds
to $p=\rho\sim a^{-6}$. But this term becomes small quickly in comparison
to radiation which goes as $\sim a^{-4}$ and a period of radiation
domination in the universe ensues. The second term in Eq.\eqref{eq:21}
gains prominence in the epoch of matter domination and we identify
it with dark matter. But as the universe evolves towards the present
era the first term begins to dominate and acts like a cosmological
constant giving rise to the late time acceleration of the universe.
The equation of state parameter after inflation is over is given by
\begin{equation}
w=\dfrac{\dfrac{k^{2}}{Ka^{6}}-m_{Pl}^{4}\left(\dfrac{L^{2}}{4K}-M\right)}{m_{Pl}^{4}\left(\dfrac{L^{2}}{4K}-M\right)+m_{Pl}^{2}\dfrac{kL}{Ka^{3}}+\dfrac{k^{2}}{Ka^{6}}}\label{eq:22}\end{equation}
 with the following values of \textit{w} corresponding to the various
epochs:

\begin{flushleft}
$w\approx1$\ \ \ \ \ \ \ \ ~~~~~~ after the end of
inflation and \\
 \ \ \ \ \ \ \ \ \ \ \ \ ~~~~~~~~\ \ ~before
radiation domination 
\par\end{flushleft}

\noindent \begin{flushleft}
$w\approx0$ \ \ \ \ \ \ \ ~~~~~~ during matter domination 
\par\end{flushleft}

\noindent \begin{flushleft}
$w\rightarrow-1$ \ \ \ \ \ \ ~~~~ as $a\mathrm{\;}\rightarrow\mathrm{\;}\infty$ 
\par\end{flushleft}

Using Eq.\eqref{eq:11} the  sound speed is found to be\begin{equation}
c_{s}^{2}=\dfrac{1}{m_{Pl}^{2}\dfrac{La^{3}}{2k}+1}\label{eq:23}\end{equation}
 From the above equation it is clear that the sound speed decreases
as the universe expands.

\section{Observational constraints}

So far we have seen that the model considered by us produces the primary
features of \textit{k}-inflation in the early universe and reproduces
dark matter as well as a cosmological constant in the later period
of evolution. We will now use various observational features to constrain
the parameters of our model. A notable feature \cite{picon,chiba}
in our model is that the potential and the kinetic part are coupled.
So parameters that are relevant during the late time era cannot be
determined independently of the parameters relevant during the inflationary
era. It is thus practical to first carry out the analysis in the late
time era and then use the calculated values of the relevant parameters
in the inflationary era. We have provided the expression for the \textit{k}-essence
energy density after inflation is over in Eq.\eqref{eq:21}. Using
the current observed value of the cosmological constant, we get \begin{equation}
m_{Pl}^{4}\left(\dfrac{L^{2}}{4K}-M\right)\simeq10^{-48}\left(GeV\right)^{4}\label{eq:24}\end{equation}

Also, observations put the current dark matter density to be about
$1/3$rd of the current dark energy density. This enables us to write
\begin{equation}
\dfrac{kL}{Ka_{0}^{3}}\approx\dfrac{1}{3}m_{Pl}^{2}\left(\dfrac{L^{2}}{4K}-M\right)\label{eq:25}\end{equation}
 where the subscript `$0$' signifies the present epoch. We know from
observations that the fraction of the current energy density contained
in radiation is $\left(\Omega_{R}\right)_{0}\simeq5\times10^{-5}$
corresponding to the present radiation density $\left(\rho_{R}\right)_{0}\simeq6.94\times10^{-53}\left(GeV\right)^{4}$.
Denoting the third term in Eq.\eqref{eq:21} as $\rho_{k}$, and assuming
that $\rho_{R}$ crosses over $\rho_{k}$ at a redshift of $z\sim10^{12}$
(prior to the nucleosynthesis at a redshift of $10^{10}$), we get
\begin{equation}
z^{2}=\dfrac{\left(\rho_{R}\right)_{0}Ka_{0}^{6}}{k^{2}}\Rightarrow\dfrac{k}{a_{0}^{3}}=\dfrac{K^{1/2}}{z}\left(\rho_{R}\right)_{0}^{1/2}\label{eq:26}\end{equation}
 Now from Eq.\eqref{eq:25} and Eq.\eqref{eq:26} we get \begin{equation}
m_{Pl}^{2}\dfrac{L}{K^{1/2}}\simeq4\times10^{-11}\left(GeV\right)^{-2}\label{eq:27}\end{equation}
 From Eqs.\eqref{eq:26} and \eqref{eq:27} it can be seen that the
the cross-over between dark matter and $\rho_{k}$ occurs at a redshift
of $\sim10^{9}$ and that between radiation and dark matter at a redshift
of $\sim10^{4}$, i.e., at the epoch of matter-radiation equality.
We also find the present value of $\rho_{k}$ to be \begin{equation}
\left(\rho_{k}\right)_{0}=\dfrac{k^{2}}{Ka_{0}^{6}}=\dfrac{\left(\rho_{R}\right)_{0}}{z^{2}}\approx6.94\times10^{-77}\left(GeV\right)^{4}\label{eq:28}\end{equation}
 The  sound speed at the epoch of matter radiation equality
turns out to be \begin{equation}
\left(c_{s}^{2}\right)_{eq}=\dfrac{1}{m_{Pl}^{2}\dfrac{La_{eq}^{3}}{2k}+1}=\dfrac{1}{m_{Pl}^{2}\dfrac{La_{0}^{3}}{2z_{eq}^{3}k}+1}\simeq4.1\times10^{-16}\label{eq:29}\end{equation}

Now, we can rexpress \textit{w} from Eq.\eqref{eq:22} in terms of
the redshift \textit{z}. Since $\rho_{k}$ is negligible in comparison
to the other components, we have \begin{equation}
w\approx\dfrac{-\left(\dfrac{L^{2}}{4K}-M\right)}{\left(\dfrac{L^{2}}{4K}-M\right)+m_{Pl}^{-2}\dfrac{kL}{Ka_{0}^{3}}(z+1)^{3}}\label{eq:29.1}\end{equation}
 Therafter, it is possible to find $dw/dz$. Its value at the current
epoch, i.e., at redshift $z=0$ using Eqs.\eqref{eq:24} and \eqref{eq:25}
turns out to be \begin{equation}
\left(\dfrac{dw}{dz}\right)_{0}\thickapprox2.733\times10^{-28}\end{equation}
 One can also estimate the current value of the equation of state
parameter in our model, which using \eqref{eq:29.1} and putting $z=0$
turns out to be \begin{equation}
w_{0}\approx-0.75\end{equation}
 We can further find out the value of the redshift at which the universe
started its transition from the matter dominated decelerating era
to its presently accelerating era. Knowing that for acceleration to
begin we must have $w=-1/3$, from Eq.\eqref{eq:29.1} we find that
\begin{equation}
z_{acc}\approx0.817\end{equation}
 Such a value for the redshift is quite compatible with present observations
\cite{Melchiorri}. But, from Eqs.\eqref{eq:24} and \eqref{eq:27}
we find that \begin{equation}
m_{Pl}^{4}M=4\times10^{-22}-10^{-48}\,(GeV)^{4}\end{equation}
 showing that a tuning of the parameter \textit{M} is needed. This
is expected since it is simply a rephrasal of the coincidence problem
associated with the present window of acceleration of the universe.

We now revisit the inflationary era for analyzing the observational
constraints pertaining to it. From Ref.\cite{Garriga} the spectrum
of scalar density perturbations in \textit{k}-inflation is given by
\begin{equation}
\begin{array}{c}
P=\dfrac{16}{9}\dfrac{m_{Pl}^{-4}}{c_{s}}\dfrac{\rho}{1+p/\rho}=-\dfrac{16}{9}\dfrac{m_{Pl}^{-4}}{c_{s}}\sqrt{\dfrac{8\pi G}{3}}\dfrac{\rho^{5/2}}{\dot{\rho}}\\
\\=\dfrac{32\sqrt{2}}{3\sqrt{3}}\dfrac{\sqrt{\pi}m_{Pl}^{-1}}{c_{s}}\dfrac{\sqrt{2}K\phi_{c}}{L}\left(\dfrac{L^{2}}{4K}-M\right)^{3/2}V_{i}^{3/2}\end{array}\label{eq:30}\end{equation}
 where in the second step we have used the energy conservation law
and also used the Friedmann equation. Using the COBE normalization
$\sqrt{P}\sim2\times10^{-5}$, and assuming that 60 e-folds of expansion
takes place, we can rewrite Eq.\eqref{eq:30} to get an expression
for $V_{i}$ to be \begin{equation}
\sqrt{V_{i}}=\dfrac{\left(27\right)^{1/6}}{4}\dfrac{c_{s}^{1/3}m_{Pl}^{-1/3}}{\pi^{1/6}}\left(\dfrac{PL}{K\phi_{c}}\right)^{1/3}\left(\dfrac{L^{2}}{4K}-M\right)^{-1/2}\label{eq:31}\end{equation}
 Using Eqs.\eqref{eq:31} and \eqref{eq:19} in Eq.\eqref{eq:17}
we can write \begin{equation}
\begin{array}{c}
c_{s}^{1/3}\phi_{c}^{2/3}=\dfrac{4}{\left(27\right)^{1/6}}\pi^{1/6}m_{Pl}^{2/3}\left(\dfrac{K}{PL}\right)^{1/3}\\
\\\left[\dfrac{1}{\sqrt{3\pi}}\dfrac{1}{L}\left(\dfrac{L^{2}}{4K}-M\right)+\dfrac{1}{\sqrt{3\pi}}\dfrac{L}{4K}+\dfrac{NL}{2^{3/2}K}\sqrt{\dfrac{3}{8\pi}}\right]\end{array}\label{eq:32}\end{equation}

Now from Eq.\eqref{eq:11} we see that in slow-roll approximation
when $F_{X}=0$ we get $c_{s}^{2}=0$. But, in the post slow-roll
stage, $X=X_{0}+\delta X$, and $F_{X}$ does not vanish. To first
order in $\delta X$ we can write $F_{X}\approx\left(F_{XX}\right)_{0}\delta X$.
Using this in Eq.\eqref{eq:11} we get\begin{equation}
c_{s}^{2}\simeq\dfrac{\delta X}{2X_{0}}\label{eq:33}\end{equation}
 Stability requires $\delta X>0$ and we show now that this is indeed
the case. From Eqs.\eqref{eq:31} and \eqref{eq:18} we calculate
$\delta X/X_{0}$ when $V=V_{i}$, to get \begin{equation}
\dfrac{\delta X}{X_{0}}=\dfrac{\frac{4K^{2}}{L^{2}}\left(\frac{L^{2}}{4K}-M\right)}{\sqrt{3\pi}\frac{4K^{2}}{L}\left[\frac{1}{\sqrt{3\pi}}\frac{1}{L}\left(\frac{L^{2}}{4K}-M\right)+\frac{1}{\sqrt{3\pi}}\frac{L}{4K}+\frac{NL}{2^{3/2}K}\sqrt{\frac{3}{8\pi}}\right]-K}\end{equation}
 It is to be noted that in order to evaluate the above equation the
actual value of \textit{K} or \textit{L} is not required, instead
the ratio $L/\sqrt{K}$ from Eq.\eqref{eq:27} serves the purpose.
Substituting the various values we find that \begin{equation}
\dfrac{\delta X}{X_{0}}\simeq2.748\times10^{-29}\label{eq:35}\end{equation}
 which is positive as claimed. The sound speed is therefore found
to be \begin{equation}
c_{s}^{2}\simeq\dfrac{1}{2}\dfrac{\delta X}{X_{0}}\simeq1.374\times10^{-29}\label{eq:36}\end{equation}
 Having found the sound speed and using the values of $P$ and $N$,
we now use Eq.\eqref{eq:27} in Eq.\eqref{eq:32} to get \begin{equation}
\dfrac{1}{\phi_{c}\sqrt{K}}\simeq3.23\times10^{15}(GeV)^{-1}\label{eq:37}\end{equation}

We now have all the parameter values to evaluate the value of \textit{V}
at the beginning and at the end of \textit{k}-inflation which we write
below \begin{equation}
V_{i}\simeq9.166\times10^{97}\label{eq:38}\end{equation}
 \begin{equation}
V_{e}\simeq1.107\times10^{94}\label{eq:39}\end{equation}
 The corresponding energy densities are \begin{equation}
\rho_{i}=V_{i}\left(\dfrac{L^{2}}{4K}-M\right)m_{Pl}^{4}\simeq9.166\times10^{49}(GeV)^{4}\label{eq:40}\end{equation}
 \begin{equation}
\rho_{e}=V_{e}\left(\dfrac{L^{2}}{4K}-M\right)m_{Pl}^{4}\simeq1.107\times10^{46}(GeV)^{4}\label{eq:41}\end{equation}

The tensor-to-scalar ratio is given by \cite{Garriga}\begin{equation}
\begin{array}{c}
r=24c_{s}\left(1+\dfrac{p}{\rho}\right)=-\dfrac{24c_{s}m_{Pl}}{\sqrt{24\pi}}\dfrac{\dot{\rho}}{\rho^{3/2}}\\
\\=\sqrt{\dfrac{24}{\pi}}\dfrac{c_{s}}{\phi_{c}}\left(\dfrac{L^{2}}{4K}-M\right)^{-1/2}\dfrac{L}{\sqrt{2}K}\dfrac{1}{\sqrt{V_{i}}}\end{array}\label{eq:42}\end{equation}
 where in the second step we have used the energy conservation and
Friedmann's equation. On substituting the parameter values we get
\begin{equation}
r=9.776\times10^{-16}\label{eq:43}\end{equation}
 The scalar spectral index can be obtained from the relation \cite{Garriga}\begin{equation}
\begin{array}{c}
n_{s}-1=-3\left(1+\dfrac{p}{\rho}\right)-\dfrac{1}{H}\dfrac{d}{dt}\ln\left(1+\dfrac{p}{\rho}\right)\\
\\-\dfrac{1}{H}\dfrac{d}{dt}\ln c_{s\,\,\,\,\,\,\,\,\,\,\,\,\,\,}\\
\\=\dfrac{2\dot{\rho}}{\rho H}-\dfrac{\ddot{\rho}}{\rho H}+\dfrac{\dot{H}}{H^{2}}-\dfrac{1}{H}\dfrac{\dot{c_{s}}}{c_{s}}\end{array}\label{eq:44}\end{equation}
 To evaluate $n_{s}$ the values of the following quantites are required\[
\dfrac{\dot{\rho}}{\rho}=\dfrac{\ddot{\rho}}{\dot{\rho}}=-\dfrac{\dot{\phi_{0}}}{\phi_{c}}=-\dfrac{m_{Pl}^{2}}{\phi_{c}}\dfrac{L}{\sqrt{2}K}=-9.136\times10^{4}GeV\]
 \[
H=\sqrt{\dfrac{8\pi G}{3}\rho_{i}}=2.771\times10^{6}GeV\]
 \[
\dot{H}=-\dfrac{4\pi G}{3}\left(\rho_{i}+p_{i}\right)=\dfrac{4\pi G}{9}\dfrac{\dot{\rho_{i}}}{H}=-4.219\times10^{10}(GeV)^{2}\]
 \[
\dfrac{\dot{c}_{s}}{c_{s}}=2.519\times10^{-94}GeV\]
 All the above values have been calculated using the slow-roll approximation
pertaining to the beginning of \textit{k}-inflation. Therefore, using
these values in Eq.\eqref{eq:44} we get \begin{equation}
n_{s}=0.96514\label{eq:45}\end{equation}
 This value is quite close to what is predicted by models of potential
driven inflation. Eq.\eqref{eq:44} differs from the appropriate expression
in the case of usual inflation by the term proportional to the derivative
of the sound speed. Since in standard inflation $c_{s}=1$, this term
vanishes and one obtains $n_{s}$ to be very close to $1$, i.e.,
a scale invariant spectrum. But in \textit{k}-inflation models, $c_{s}\neq1$,
and a tilted spectrum with $n_{s}<1$ is generally predicted. However,
in our model this term in Eq.\eqref{eq:44} makes a vanishingly small
contribution, and hence we get a spectral index that is again quite
close to $1$. Only the value of the tensor-to-scalar ratio in our
model makes it distinguishable from standard inflation where typically
a value of about 0.12 to 0.15 is obtained.

Now, the duration of inflation in our model is found to be \begin{equation}
t_{e}-t_{i}=\intop_{\phi_{i}}^{\phi_{e}}\dfrac{d\phi}{\dot{\phi}}=\dfrac{\sqrt{2}K}{L}\left(\phi_{e}-\phi_{i}\right)m_{Pl}^{-1}\approx6.9\times10^{-29}s\label{eq:46}\end{equation}

After the end of inflation, the stage of kinetic dominated evolution
sets in very quickly. In order to have an idea as to how much time it takes 
for the exponential part of the potential to become
negligible, we assume that for argument's sake, $X\simeq X_{0}$.
This assumption is only made to perform a simple calculation and get
an upper bound on the time required for the exponential part to decay
(the actual time taken is much smaller since $X>X_0$ and $\phi$ evolves
more rapidly compared to its linear evolution during inflation).
The time taken after inflation for the exponential part to attain the value
$e^{-\phi/\phi_{c}}\simeq0.01$, is about $1.7\times10^{-27}s$.
Thus, the time required for the \textit{k}-essence field to effectively
behave as kinetic \textit{k}-esence is of the order of $10^{-27}s$.
This again justifies our analysis of the previous section pertaining to
the post inflationary period being dominated by the dynamics of purely
kinetic $k$-essence. It should be noted that the estimate for the
time required for the universe to enter into a kinetic dominated era
after inflation is actually an upper bound. In reality the time required
is much shorter since $X>X_{0}$ and the scalar field evolves more
rapidly with time than during the inflationary era (the potential
decreases very quickly to assume an almost constant value).

Reheating in this model could be caused by gravitational particle
production. The process of gravitational reheating in the presence
of kinetic domination by a scalar field is not yet understood very
well \cite{Chun}. However, standard calculations \cite{Ford} give
the density of particles produced at the end of inflation to be $\rho_{R}\simeq8.67\times10^{15}g(GeV)^{4}$
where \textit{g} is the number of fields which produce particles at
this stage, likely to be between 10 and 100. This energy density if
immediately thermalized would give rise to a temperature of $T_{e}\simeq9.65\times10^{3}\left(\dfrac{g}{g_{*}}\right)^{1/4}GeV$,
where $g_{*}$ is the total number of species in the thermal bath
and maybe somewhat higher than \textit{g}. Assuming that immediately
after the end of inflation there is complete kinetic domination so
that the scalar field density falls as $a^{-6}$, it is estimated
that in our model the universe has to expand by a factor of about
$10^{15}$ for radiation domination to set in. After that expansion
the temperature which goes as $T\propto1/a$ comes out as $T\simeq9.65\times10^{-12}\left(\dfrac{g}{g_{*}}\right)^{14}GeV$.
So we see that the temperature is not high enough for a successful
nucleosynthesis for which a temperature around 1 MeV is needed. Now,
if we change our parameters somewhat such that the value of the redshift
for the cross-over between $\rho_{R}$ and $\rho_{k}$ is $10^{6}$,
then we find that the reheat temperature turns out to be $T\simeq9.65\times10^{-5}\left(\dfrac{g}{g_{*}}\right)^{1/4}GeV$
which is roughly about the order of $0.1$ MeV. There have been some
recent studies which indicate that very low reheating temperatures
could also be a viable option for successful nucleosynthesis (see,
for instance \cite{Kohri}). These ideas have to be analyzed in detail
in the context of $k$-essence scenarios in order to check how far
gravitational reheating could be successful in our model.

\section{Conclusions}

To summarize we have considered a \textit{k}-essence model that produces
inflationary expansion in the early universe by the process 
of \textit{k}-inflation
and later on generates both dark matter and dark energy at appropriate
subsequent stages. For our Lagrangian we have considered that form
which has been widely used for \textit{k}-essence models \cite{chiba}.
In contrast to an earlier model studied by us \cite{Bose}, the potential
and the kinetic parts of the scalar field are not decoupled, leading
to coupling between the inflationary era and the late time parameters.
A significant feature of this fact can be found in the expression
for the energy density. It thus follows that the generated cosmological
constant which dominates the dynamics at late times, derives its value
from inflationary parameters. It needs to be mentioned here that our model
is unable to address the coincidence problem. The addressal of this problem
within the context of \textit{k}-essence is made possible by the existence 
of fixed points in the radiation and matter era.
In order to have these fixed points, it is necessary that the potential  has
the form $V(\phi)=1/\phi^2$. It was shown that such models
that solve the coincidence problem suffer from superluminal propagation of the
field perturbations \cite{superlum}  (which, however, may not 
affect causality \cite{causal}). But the 
choice of the potential in our model does not allow the
existence of fixed point in the radiation and matter era. Consequently,
this model does not suffer from the problem of superluminal propagation.

Our model is able to reproduce the basic features of \textit{k}-inflation.
Although in general \textit{k}-inflation predicts that $n_{s}<1$,
our model gives rise to a value which is nearly the same with what
is obtained in standard potential diven inflation, predicting an almost
scale invariant density perturbation spectrum. But, the value of the
calculated tensor-to-scalar ratio is quite different from what is
obtained in standard inflationary models. 
After the inflation is over
the potential quickly becomes constant and we are able to approximate
the model as purely kinetic \textit{k}-essence. The late time energy
density and the sound speed in terms of the scale factor $a$ were
obtained. The resultant energy density contained terms that achieved
the desired unification of dark matter and dark energy. We showed
that the sound speed calculated at the epoch of matter-radiation equality
came out to be very small, thus posing no problem for structure formation,
since it further decreases as the universe expands. Our estimation
of the current equation of state and the redshift at which the current
acceleration of the universe started, lie within observational bounds.
Further studies would be needed to see if gravitational reheating
could be a viable feature of such a scheme.

{\it Acknowledgements}: ASM would like to acknowledge support from the
DST project SR/S2/PU-16/2007.

\end{document}